\documentclass[12pt]{article}

\usepackage[dvips]{color}

\renewcommand{\title}[1]{\begin{center}\bf\Large #1\end{center}}

\renewcommand{\author}[1]{\begin{center}\large #1\end{center}}

\usepackage{latexsym,amsfonts}
\newcommand{\rr}{\mathbb{R}}

\begin{document}
\begin{titlepage}
\hspace{12.6cm}{\bf HU-EP-05/36}

\vspace{5mm}

\title{Massless scalar particle on AdS spacetime:\\ 
Hamiltonian reduction and quantization}

\vspace{5mm}

\author{

Harald Dorn ${}^a$
  and George Jorjadze ${}^b$\\
{\small${}^a$Institut f\"ur Physik der
Humboldt-Universit\"at zu Berlin,}\\
{\small Newtonstra{\ss}e 15, D-12489 Berlin, Germany}\\
{\small${}^b$Razmadze Mathematical Institute,}\\
{\small M.Aleksidze 1, 0193, Tbilisi, Georgia}}

\vspace{5mm}

\begin{abstract}
\noindent
We investigate the massless scalar particle dynamics on $AdS_{N+1}~ (N>1)$
by the method of Hamiltonian reduction. Using the dynamical 
integrals of the conformal symmetry we construct the physical 
phase space of the system as a $SO(2,N+1)$ orbit in the space of 
symmetry generators. The symmetry generators themselves are
represented in terms of $(N+1)$-dimensional oscillator variables.
The physical phase space establishes a correspondence 
between the $AdS_{N+1}$ null-geodesics
and the dynamics at the boundary of $AdS_{N+2}$.
The quantum theory is described by a UIR of $SO(2,N+1)$ obtained
at the unitarity bound. This representation contains a pair of UIR's 
of the isometry subgroup $SO(2,N)$ with the Casimir number corresponding to 
the Weyl invariant mass value. The whole discussion includes the globally
  well-defined realization of the conformal group via the conformal embedding
of $AdS_{N+1}$ in the ESU $\rr\times S^N$.

\vspace{5mm}



\end{abstract}

\end{titlepage}
\baselineskip=20pt

\section{Introduction}

Starting already in the sixties, there exists an extensive literature on 
unitary irreducible representations
(UIR's) of $SO(2,N)$ and their use either for the conformal group of Minkowski
spaces or for the isometry group of $AdS_{N+1}$. In recent times the interest
in this subject was renewed in particular by the AdS/CFT correspondence
\cite{ADSCFT}. This paper is the third in a series devoted to the study of
UIR's of  $SO(2,N)$ in relation to particle dynamics on $AdS_{N+1}$.
Our results concern the explicit
realization of UIR's $D_N(\alpha)$ of $SO(2,N)$, for generic $N$, either in terms of 
operators acting
on holomorphic functions of $N$ complex variables, obtained via geometric
quantization \cite{DJ1}, or $N$-dimensional oscillator variables \cite{DJ2}.
Furthermore our study is singled out by being completely based
on particle dynamics and its treatment via Hamilton reduction.
Within this framework the present paper is devoted to the study of the
peculiarities of massless scalar particles. In the field theoretical
treatment these peculiarities have been shown to be related to singleton 
representations \cite{F1,F2,F3}.   

The classical action for a particle in $AdS_{N+1}$ contains its classical mass
$m$ and the  $AdS_{N+1}$ radius $R$ as parameters
. For generic $m\neq 0$ the symmetry group is $SO(2,N)$, as
the isometry group of the space-time. On the classical level the quadratic 
Casimir $C=\frac{1}{2}J_{AB}J^{AB}$ is equal to $m^2R^2$ and the lowest
possible particle energy $\alpha$ in units of $1/R$ is equal to $mR$.
Quantization \cite{DJ1} deforms the classical relation $C=\alpha ^2$ to
$C_q=\alpha _q(\alpha _q-N)$. $C_q/R^2$ is interpreted as the squared mass
of the quantum particle and the connection to the classical mass is lost.
\footnote{After this comment we will drop the index $q$ for the quantum
  versions.} 

Therefore, up to this point the question which value of the Casimir (mass) 
in the quantum case corresponds to the massless particle cannot be answered.
The only possibility to identify the Casimir (mass) of the massless particle
is via its enhanced symmetry related to conformal invariance. To do this,
there are at least two possibilities. At first one can take into account
information from outside particle dynamics and identify the mass of the
quantum particle with the mass in the classical Klein-Gordon equation.
There one has Weyl invariance for $C=-\frac{N^2-1}{4}$, corresponding
to $\alpha =\frac{N\pm 1}{2}~.$ We will comment this option later in the
paper, but our main concern will be connected to a second possibility.
As announced, we stay completely within particle dynamics, start with the
classical action of a massless particle, note its invariance under
the larger conformal group of $AdS_{N+1}$, which is $SO(2,N+1)$,
and quantize this symmetry group. 

Various subtleties we will meet in the following are related to
the somewhat exotic causal structure of $AdS_{N+1}$. This space-time
is not a global hyperbolic one. Null geodesics reach the boundary
in finite time. There are conjugate points, in particular 
all time-like geodesics starting at one point meet again after a time
interval $\pi$ at the corresponding $AdS$ antipodal point, and related to this,
only part of the causal future of a point can be reached by time-like
geodesics starting at this point \cite{EH,F1}. Although these issues
play a role in field theory, both for the massless as well as the massive case
via the specification of certain boundary conditions, there seems to be
no particular problem for generic masses in particle dynamics,
the particle stays within the space-time for all times and the
space of particle trajectories is mapped one to one to the space
of dynamical integrals \cite{DJ2}. However, the situation becomes
more involved for the massless particle, its classical trajectories go from
boundary to boundary within a time interval $\pi $ and 
to a given set of isometric dynamical integrals belong two trajectories.
Special care is needed
also for the global realization of the conformal group itself.

\setcounter{equation}{0}

\section{The conformal group of $AdS_{N+1}$} 

The  $(N+1)$-dimensional $AdS$ space can be realized as the hyperboloid
of radius $R$
\begin{equation}\label{hyp}
 X_0^2+X_{0'}^2-\sum_{n=1}^NX_n^2=R^2
\end{equation}
embedded in the $(N+2)$-dimensional flat space $\rr_N^2$
with coordinates $X_A,$ $A=(0,0',1,...,N)$
and the metric tensor $\,G_{AB}=\mbox{diag}(+,+,-,...,-)$.
\footnote{More precise $AdS_{N+1}$ is understood as the universal covering of the
  hyperboloid. If below we talk about its  isometry group, we 
  have in mind the related universal covering of $SO(2,N)$.}

As is well-known, the conformal group of the four-dimensional Minkowski space,
due to the fact that the special conformal transformations map certain
whole light
cones to infinity, is globally not well defined within the original space. 
To cure
this problem one has to enlarge the discussion to an infinite-sheeted
covering of the Minkowski space, which can be conformally mapped to an 
Einstein static universe (ESU) \cite{LM}. It is straightforward to adopt
the techniques in this construction to our situation.

From general theorems on constant curvature spaces it is known that the 
conformal group of $AdS_{N+1}$ for
$N\geq 2$ is $SO(2,N+1)$, the same as for the Minkowski space of the same
dimension. The case $N=1$ has infinite dimensional conformal symmetry
and will not be discussed here. What we need are explicit transformation formulas. To derive
them we start with the cone in $\rr_{N+1}^2$
\begin{equation}\label{cone}  
Y_0^2+Y_{0'}^2-Y_1^2-\dots -Y_{N}^2-Y_{N+1}^2~=~0~
\end{equation}
without its vertex $Y=0$. It is invariant under $SO(2,N+1)$ transformations 
\begin{equation}\label{linreal}
Y^{\hat A}~\mapsto ~\Lambda ^{\hat A}_{~\hat B}~Y^{\hat B}~,~~~~~~\hat A,~\hat
B~=~0,0',1,\dots N,N+1~.
\end{equation}
We now relate the cone (\ref{cone}) to the  $AdS$ hyperboloid (\ref{hyp}) via
\begin{equation}\label{AdS-cone}
X^A~=~R~\frac{Y^A}{Y^{N+1}}~,~~~Y^{N+1}>0~,~~~A~=~0,0',1,\dots N~
\end{equation}
and get under a $SO(2,N+1)$ transformation
\begin{equation}\label{AdSconf}
X^A~\mapsto ~\frac{\Lambda ^A_{~B}~X^B~+~R~\Lambda ^A_{~N+1}}{\Lambda
  ^{N+1}_{~~N+1}~+~R^{-1}\Lambda ^{N+1}_{~~~\,B}~X^B}~.
\end{equation}
It is straightforward to check that these maps of the hyperboloid are indeed conformal ones.

Since $\Lambda \in SO(2,N+1)$, one has 
$1=(\Lambda ^{N+1}_{~~N+1})^2-\Lambda^{N+1}_{~~~\,A}\Lambda ^{N+1~A}$. 
Therefore, either $\Lambda^{N+1}_{~~~\,A}=0,~\forall A$ and
$\Lambda^{N+1}_{~~~N+1}=\pm 1$, 
or $\Lambda^{N+1}_{~~~\,A} $ is a non-zero
vector in the embedding space of the hyperboloid. In the first case
the transformation (\ref{AdSconf}) is an isometry of the $AdS$ hyperboloid
and globally well defined. In $all$ other cases the transformation is
singular where the denominator vanishes. This means
that isometries are the only globally well defined
conformal transformations.

We now map the cone (\ref{cone}) to an ESU $\rr\times S^N$, defined
by $\theta \in \rr $, $(\vec Z,Z^{N+1})\in \rr ^{N+1}$, $\vec
Z^2+(Z^{N+1})^2=1$ via
\begin{equation}\label{coneESU} 
\tan\theta ~=~
\frac{Y^{0'}}{Y_0}~,~~~\vec Z~=~\frac{\vec Y}{\sqrt{Y_0^2+Y_{0'}^2}}~,~~~
~Z^{N+1}~=~\frac{Y^{N+1}}{\sqrt{Y_0^2+Y_{0'}^2}}~.
\end{equation}
Due to (\ref{AdS-cone}) this induces the standard injective map of $AdS_{N+1}$
to the ESU
\begin{equation}\label{AdSESU} 
\tan\theta ~=~
\frac{X^{0'}}{X_0}~,~~~\vec Z~=~\frac{\vec X}{\sqrt{X_0^2+X_{0'}^2}}~,~~~
~Z^{N+1}~=~\frac{R}{\sqrt{X_0^2+X_{0'}^2}}~.
\end{equation}
Since in the last formula  $0<Z^{N+1}\leq 1$, the image of $AdS_{N+1}$ is 
just a half of the ESU. The other half of ESU then can be considered as a
copy of the original  $AdS_{N+1}$. The conformal boundary of $AdS_{N+1}$ is 
mapped
to $\rr$ times the $(N-1)$-dimensional equator at $Z^{N+1}=0$ of the
$N$-dimensional ESU sphere. 
If one introduces on the cone (\ref{cone}) equivalence classes $Y^{\hat A}\sim
\mu Y^{\hat A},~~\mu >0$ the ESU is mapped one to one to these equivalence
classes. Since the action of $SO(2,N+1)$ commutes with this equivalence
relation, it is globally well defined on the ESU. 

A last comment concerns
the differences to the Minkowski case  \cite{LM}. There its conformal
boundary is fixed by $Y^{0'}+Y^{N+1}=0$. Due to this 
an infinite number of conformal copies of Minkowski space find its place on
the  ESU. Furthermore, there at least some of the $SO(2,N+1)$ transformations, 
mixing $Y^{N+1}$ with other coordinates, keep the boundary invariant. They are
combinations of isometries and dilatations.

\setcounter{equation}{0}

\section{Classical theory}

A part of the classical description for massless and massive particle 
dynamics is similar. Therefore, we briefly repeat the scheme of \cite{DJ1}
for $m=0$ and  discuss the items specific for the massless case
in more detail.

The massless particle dynamics on
the hyperboloid (\ref{hyp}) is described by the action 
\begin{equation}\label{S}
S=-\int d\tau \left[\frac{\dot X^A\dot X_A}{2e}+
\frac{\mu}{2}\left(X^AX_A-R^2\right)\right]~,
\end{equation}
where $e$ and $\mu$ are Lagrange multipliers and $\tau$ is an
evolution parameter. 
The role of the time coordinate on the hyperboloid (\ref{hyp})
is played by the polar angle $\theta$ on the  plane $(X_0, X_{0'})$:
$X_{0}=r\cos\theta$, $X_{0'}=r\sin\theta$.
To have the kinetic term of the space coordinates $\dot X_n\dot X_n$ 
with a positive coefficient we assume $e>0$, and to fix the time 
direction we choose $\dot\theta>0$, 
which is equivalent to $ {X_0\dot X_{0'}-X_{0'}\dot X_0}>0$.

The action (\ref{S}) is gauge invariant and by the Dirac procedure 
we
find three constraints
\begin{equation}\label{Phi=0}
X^AX_A-R^2=0~,~~~~~~~~P_AP^A=0~,~~~~~~~~~ P_A\,X^A=0~.
\end{equation}
The constraint $P_AP^A=0$ is of the first class, whereas the
two others are of the second class. 
Therefore, the dimension of the reduced phase
space is $2N$ like in the massive case. 
We will describe this space in terms of dynamical integrals.

The spacetime isometry group $SO(2,N)$ 
provides the  conserved quantities
\begin{equation}\label{M_AB}
J_{AB} =P_A\,X_B -P_B\,X_A~,
\end{equation}
where $P_A$ are the canonical momenta $P_A= -{\dot X_A}/e$.
Since $\theta$ is the time
coordinate, $J_{00'}$ is associated with the particle energy $E$
and due to our assumptions it is positive
\begin{equation}\label{E}
E=P_0X_{0'}-P_{0'}X_0= e^{-1}(X_0\dot X_{0'}-X_{0'}\dot X_0)>0~.
\end{equation}
The  boosts we denote by $J_{0n}=K_n$, $J_{0'n}=L_n$ and
we also use their complex combinations 
\begin{equation}\label{z_n}
z_n=L_{n}-iK_{n}~,~~~~~~~~z_n^*=L_{n}+iK_{n}~,~~~n=1,\dots ,N~.
\end{equation}
For further calculations it is convenient to
introduce the following $SO(N)$ scalars
\begin{equation}\label{z_nz_n}
J^2=\frac{1}{2}\,
J_{kk'}J_{kk'}~,~~~~~~\lambda^2=z^*_kz_k~,~~~~~
\rho^2= \sqrt{{z^*}^2\,z^2}~,~~~~~~
e^{2i\beta}= \frac{z^2}{\rho^2}~, 
\end{equation}
where $z^2= z_kz_k$, ${z^*}^2= {z^*_kz^*_k}\,$ and we assume $0\leq\beta<\pi$.

Due to the constraints (\ref{Phi=0}) the $SO(2,N)$  quadratic Casimir number 
vanishes, $C=\frac{1}{2}\, J_{AB}J^{AB}=0$, and this condition can be written as
\begin{equation}\label{C1}
E^2+J^2=\lambda^2~.
\end{equation}
A set of other quadratic relations 
$~J_{AB}\,J_{A'B'}= J_{AA'}\,J_{BB'}-J_{AB'}\,J_{BA'}~$
follows from (\ref{M_AB}) as
identities in the variables $(P, X)$.
Taking $A=0$, $B=0'$, $A'=m$, $B'=n$ $(m\neq n)$ and using
(\ref{z_n})  we obtain
\begin{equation}\label{EM=zz}
2iE\,J_{mn}=z_m^*z_n-z_n^*z_m~.
\end{equation}
Its square yields $4E^2J^2=\lambda^4-\rho^4$ and together with (\ref{C1}) 
we find the following relations between the scalar variables
\begin{eqnarray}\label{E,J_0}
E^2=\frac{1}{2}\,\left(\lambda^2+\rho^2\right)~,~~~~~~~~
J^2=\frac{1}{2}\,\left(\lambda^2-\rho^2\right)~.
\end{eqnarray}
 
Eqs.  (\ref{z_nz_n}), (\ref{EM=zz}), (\ref{E,J_0})  
define $E$ and $J_{mn}$ as functions
of ($z_n, z_n^*$) and, therefore, ($z_n, z_n^*$) or ($K_n, L_n$) 
are global coordinates on the space of dynamical integrals of the isometry
group. These integrals allow to represent the particle 
trajectories geometrically without solving the dynamical
equations. From (\ref{M_AB}) we find $N$ equations as identities in
the variables $(P,X)$
\begin{equation}\label{traj}
E\,X_n=K_{n}\,X_{0'}-L_{n}\,X_0~.
\end{equation}
Since $E$, $K_{n}$, $L_{n}$ are constants, Eq. (\ref{traj})
defines a 2-dimensional plane in the embedding space $\rr_N^2$.
The intersection of this plane with the hyperbola (\ref{hyp}) is a
particle trajectory. The plane defined by (\ref{traj}) goes through the
origin of $\rr_N^2$ and the way how it intersects the hyperboloid
depends on the relations between the dynamical integrals $K_n$, $L_n$ and 
$E$. To describe the character of trajectories (\ref{traj}) we
parameterize them by the time coordinate $\theta$
\begin{eqnarray}\label{traj1}
&&X_0=r(\theta)\cos\theta~,~~~~ X_{0'}=r(\theta)\sin\theta~,
~~~~X_n=\frac{r(\theta)}{E}\left(K_n\,\sin\theta-
L_n\,\cos\theta\right)~,\\ \label{r_0}
&&~~~~~~~~~~~\mbox{with}~~~~~~~~~~
r(\theta)=\frac{ER} {\rho\,|\sin(\theta-\beta)|}\,\,~.
\end{eqnarray}
The function $r(\theta)$ here is obtained from the relation
$X_nX_n=r^2-R^2$ and Eqs. (\ref{z_nz_n}), (\ref{E,J_0}).

The singularities of $r(\theta)$ correspond to the
$AdS$ boundary and therefore for $\rho=0$ the massless
particle is always at the boundary. From the isometric point of view
this would force us to remove $\rho =0$ out of the phase space. However,
for implementing conformal invariance one anyway has to switch
to the ESU whose one half is conformally
mapped to the $AdS$. Then $\rho =0$ has to be kept within the phase
space. The corresponding trajectories are completely inside the equator
of the ESU.

If $\rho\neq 0$,
the singularity of (\ref{r_0}) at $\theta-\beta=k\pi$ indicates
that the massless particle always reaches the $AdS$ boundary
and, for a fixed $(K_n,L_n)$, there are two different null-geodesics
given in the time intervals $\theta\in
(\beta,\pi+\beta)|_{\mbox{\scriptsize mod}\,2\pi}$ and
$\theta\in (\pi+\beta,2\pi+\beta)|_{\mbox{\scriptsize mod}\,2\pi}$,
respectively. The pieces of both trajectories, which are disconnected with
respect to $AdS$, represent the ``visible'' parts of two smooth trajectories
in the full ESU.  These are just two with luminal velocity driven great 
circles on the ESU sphere, which intersect each other on the equator.

To complete the description of the reduced phase space and its relation
to the $AdS$ null-geodesics we introduce 
additional dynamical integrals related to the conformal symmetry. 
The action (\ref{S}) is invariant under the conformal transformations, 
since the conformal factor of the kinetic
term can be compensated by the transformations of the
Lagrange multiplier $e$. 
Considering infinitesimal transformations (\ref{AdSconf}) 
different from isometries, we find the Killing vectors (the down index
labels them) 
\begin{equation}\label{Killing}
{\cal K}^B_{A}=R\,\delta_{A}^B-R^{-1}\,X_A\,X^B~.
\end{equation}
The corresponding conserved
quantities are $C_A={\cal
K}^{B}\,_A\,P_{B}=RP_A-R^{-1}(PX)\,X_A$ and on the constraint
surface (\ref{Phi=0}) they become
\begin{equation}\label{C_A}
C_A=RP_A~.
\end{equation}
The conservation of canonical momenta $P_A$ in the
massless case can also be checked directly from the dynamical
equations $~\dot X_A=-e P_A$, $~\dot P_A=\mu X_A$.
Multiplying the first equation by $P^A$, the
second by $X^A$ and using the constraints (\ref{Phi=0}),
we find $\mu=0$, which provides $\dot P_A=0$. Due to the conservation
of  $P_A$, the null-geodesics are straight lines in  $\rr_N^2$.
Making use of (\ref{traj1}) and (\ref{r_0})
we represent the trajectories as $X_A=C_A\,T+Q_A,$
where $T={ER}\,\rho^{-2}\,\,\cot(\theta-\beta)$ is a parameter
along the lines, and 
\begin{eqnarray}\label{Q_A}
Q_0=\mp\,\frac{ER}{\rho}\,\,\sin\beta~,~~~
Q_{0'}=\pm\,\frac{ER}{\rho}\,\,\cos\beta~,~~~
Q_n=\pm\,\frac{R}{\rho}\,(K_n\cos\beta-L_n\sin\beta)\,.
\end{eqnarray}
The two signs above correspond to two lines given for the same
set of isometry generators.

Since the number of functionally independent dynamical integrals has to be 
$2N$, we investigate relations between 
$J_{AB}$ and $C_A$. From  (\ref{M_AB}) and (\ref{C_A}) we find
\begin{equation}\label{EC=KC-LC}
E\,C_n=K_n\,C_{0'}-L_n\,C_0~~~~~~~~~~~(n=1,...,N)~,
\end{equation}
as identities in the variables  $(P,X)$, like in (\ref{traj}). Two other
relations
\begin{equation}\label{KL=CC} K_k\,L_k
+C_0\,C_{0'}=0~~~~~\mbox{and}~~~~K_k\,K_k-L_k\,L_k+C_0^2-C_{0'}^2=0~
\end{equation}
also follow from (\ref{M_AB}) and (\ref{C_A}), but on the
mass-shell (\ref{Phi=0}) only.
Eqs. (\ref{EM=zz}), (\ref{E,J_0}), (\ref{EC=KC-LC})  define all
generators of the conformal symmetry as functions of $K_n$, $L_n$,
$C_0$, $C_{0'}$ and these $2N+2$ dynamical integrals  are
constrained by (\ref{KL=CC}). Introducing the complex variable
$w=C_{0'}-iC_0$, the two equations of (\ref{KL=CC}) can be combined in
a one complex relation
\begin{equation}\label{zz+ww=0}
z^2 +w^2=0~.
\end{equation}
Because of $E>0$, the vector $z_n$ is nonzero and, therefore,
the space of dynamical
integrals defined by (\ref{zz+ww=0}) is a regular $2N$-dimensional 
manifold. This manifold is
identified with the reduced phase space, which is the physical phase space
$\Gamma_{ph}$ of the system.
The isometry generators ($K_n\,,L_n$)
are only local coordinates on $\Gamma_{ph}$ and by (\ref{zz+ww=0}) we have
\begin{equation}\label{w}
w=\mp i\sqrt{z^2}=\mp i\rho
e^{i\beta}~.
\end{equation}
Eq. (\ref{E}) yields  
$E=e^{-1} r^2(\theta)\dot\theta$ and we can express the Lagrange multiplier
$e$ through the dynamical variables. Then, calculating 
$w=Re^{-1}(\dot X_{0'}-i\dot X_{0})$ on the trajectories 
(\ref{traj1})-(\ref{r_0})
for the two different intervals $\theta\in (\beta,\pi+\beta)$ and 
$\theta\in (\pi+\beta,2\pi+\beta)$,
we find $w=-i\rho e^{i\beta}$ 
and $w=i\rho e^{i\beta}$, respectively.  Thus,
the above mentioned two trajectories correspond to the two possible values
of $w$ in (\ref{w}). 

As far as the velocities of the massless particle are constrained by
$\dot X_A\dot X^A=0$, the set of null-geodesics is
$(2N-1)$-dimensional and, unlike to the massive case, there is no
one to one correspondence between the trajectories and the space
of dynamical integrals. Nevertheless, the set of all trajectories
reflects the structure of $\Gamma_{ph}$. They are invariant under 
re-scalings
of all dynamical integrals $J_{AB}\mapsto e^\gamma\,J_{AB}$. 
Hence  the set of trajectories can be identified 
with $\Gamma_{ph}/\rr_{+}$.

Now we describe the Poisson bracket structure of $\Gamma_{ph}$.
The $so(2,N)$ Poisson bracket algebra of the isometry generators 
(\ref{M_AB}) is obviously preserved after the reduction to $\Gamma_{ph}$.
It can be written in the form
\begin{eqnarray}\label{PB_z}
&&\{z_m^*,\,z_n\}=2J_{mn}-2i\delta_{mn}\,E~,~~~~~
\{z_m,\,z_n\}=0=\{z_m^*,\,z_n^*\}~,\\\label{PBE,J}
&&\{J_{lm},\,z_n\}=z_l\,\delta_{mn}-z_m\,\delta_{ln}~,
~~~~~~\{E,\,z_n\}=-iz_n~,~~~~~ \{E,\,J_{mn}\}=0~,
\end{eqnarray}
and the $J_{mn}$'s form a $so(N)$ subalgebra.
Since the conformal generators $C_A$ do not commute with the second
class constraints of (\ref{Phi=0}), their Poisson brackets 
are deformed after the Hamiltonian
reduction to $\Gamma_{ph}$. To calculate such reduced brackets we use that
\begin{equation}\label{PBz^2}
\{z_m^*,z^2\}=\frac{2}{iE}\left(\rho^2\,z_m+z^2\,z_m^*\right)~,
\end{equation}
which follow from (\ref{PB_z})  due to (\ref{E,J_0}).
Then, writing (\ref{EC=KC-LC}) in the form $2iEC_n=z_n^*w-z_nw^*$,
from (\ref{zz+ww=0}) and (\ref{PBz^2}) we find
\begin{eqnarray}\label{PBw}
&&\{z_n^*,\,w\}=2C_{n}~,~~~~~~\{w,\,w^*\}=2i\,E~.
~~~~~~~\{z_n,\,w\}=0~.
\end{eqnarray}
These equations define the Poisson brackets between other conformal
generators  and  the result can be written in the form
\begin{equation}\label{PBC,J}
\{J_{AA'},C_B\}=G_{AB}\,C_{A'}-G_{A'B}\,C_{A}~,~~~~~~~
\{C_{A},C_B\}=-J_{AB}~.
\end{equation}

The Poisson brackets (\ref{PBC,J}) extend the $so(2,N)$ algebra
(\ref{PB_z})-(\ref{PBE,J}) up to $so(2,N+1)$, which describes
the underlying conformal symmetry. Adding one column
and one row to the antisymmetric matrix $J_{AB}$ by
the scheme
\begin{equation}\label{JC}
J_{\hat A\hat B}=\left(\begin{array}{cr}
J_{AB}&C_A\\-C_A&0\end{array}\right)~,~~~~~~~~\hat A, \hat
B=(0,0',1,...,N+1)
\end{equation}
we get $(N+3)\times(N+3)$ antisymmetric $ J_{\hat A\hat B}$ and
the Poisson brackets of its components correspond to the $so(2,N+1)$ 
algebra in the standard covariant form.

We specify the compact and non-compact
generators for the $SO(2,N+1)$ symmetry by
$E$, $\,J_{\hat m\hat n}$
and $z_{\hat n}$, $\,z^*_{\hat n}\,$ 
($\hat m,\hat n=1,..., N+1$), respectively. So, we use
the same notations as for $SO(2,N)$, 
only the indices run from $1$ to
$N+1$.  To distinguish the $SO(N+1)$ scalars we 
use the sign `hat'
\begin{equation}\label{tilde}
\hat J^2=\frac{1}{2}\,
J_{\hat k\hat k'}J_{\hat k\hat k'}~,~~~~~
\hat \lambda^2=z^*_{\hat k}z_{\hat k}~,~~~~
\hat z^2= z_{\hat k}z_{\hat k}~,~~~~
\hat{z^*}^2= z^*_{\hat k}z^*_{\hat k}.
\end{equation}
Due to the mass-shell condition $P_AP^A=0$ and the relation
$C_A=RP_A$,  the Casimir number
for $ J_{\hat A\hat B}$ is also zero. In terms of
$SO(N+1)$ scalars this condition reads
\begin{equation}\label{C_N+1}
E^2+\hat J^2-\hat\lambda^2=0~.
\end{equation}
Eq. (\ref{EC=KC-LC}) provides that the quadratic relations (\ref{EM=zz})
are fulfilled for the components $J_{n\,N+1}$ as well. As a result
\begin{equation}\label{EM=zz1}
2iE\,J_{\hat m\hat n}=z_{\hat m}^*z_{\hat n}-
z_{\hat n}^*z_{\hat m}
\end{equation}
is valid for all  indices $\hat m$ and $\hat n$.
Finally, the condition (\ref{zz+ww=0}) becomes
\begin{equation}\label{tildez^2}
\hat z^2=0~.
\end{equation}

Thus, the physical phase space of the massless
particle $\Gamma_{ph}$ is identified with the space of
$SO(2,N+1)$ generators $E$, $J_{\hat m\hat n}$, $z_{\hat n}$
$z^*_{\hat n}$, which satisfy the Eqs. (\ref{C_N+1})-(\ref{tildez^2})
with $E>0$ and non-zero vector $z_{\hat n}$. 
Note that, due to (\ref{PBz^2}) and the trivial Poisson brackets\\
$ \{z_m,z^2\}=0, \{J_{mn},z^2\}=0, \{E,z^2\}=-2iz^2$, the condition $z^2=0$ is
$SO(2,N)$ invariant and, therefore, the
manifold $\hat z^2=0$ is $SO(2,N+1)$ invariant.

\setcounter{equation}{0}

\section{Quantum theory}

A consistent quantum theory of the massless particle should provide 
a realization of the $SO(2,N+1)$ symmetry based of the classical picture.
The Poisson bracket relations of the $so(2,N+1)$ algebra (\ref{PBC,J}) are 
essentially non-linear in terms of the
independent variables and their direct representation 
seems more complicated than in the massive case, since one has to realize
a higher symmetry on a non-trivial phase space of the same dimensionality.

We  realize the $SO(2,N+1)$ symmetry by representations 
$D_{N+1}(\alpha)\,$ \cite{DJ2}, which are based on the creation-annihilation 
operators $(a_{\hat m}^*,~a_{\hat m})$ of a $(N+1)$-dimensional oscillator.
The generators of $SO(N+1)$ rotations in $D_{N+1}(\alpha)$
have the standard quadratic form
\begin{equation}\label{M_mn^}
J_{{\hat m}{\hat n}}=i(a^*_{\hat n} a_{\hat m}- 
a^*_{\hat m} a_{\hat n})~,
\end{equation}
and the operator $\hat J^2$ in terms of the creation-annihilation
operators, respectively, is
\begin{equation}\label{M^2^}
{\hat J}^2=\frac{1}{2} J_{{\hat m}{\hat n}} J_{{\hat m}{\hat n}}
={\hat H}^2+(N-1)\hat H- {\hat a}^*\,^2{\hat a}^2~.
\end{equation}
Here $\hat H=a^*_{\hat n} a_{\hat n}$ is the normal ordered 
$(N+1)$-dimensional
oscillator Hamiltonian,
$\hat a^2= a_{\hat n} a_{\hat n}$ and ${\hat a}^*\,^2= 
a^*_{\hat n} a^*_{\hat n}$.

The energy operator $E$ is given as a shifted oscillator Hamiltonian 
\begin{equation}\label{E^}
E=\hat H+\alpha~,
\end{equation}
and $\alpha$ coincides with its minimal eigenvalue. The compact
subalgebra $so(2)\times so(N+1)$, thus, is realized automatically.

The operators $z_{\hat n}$ and $ z^*_{\hat n}$ are represented by
\begin{eqnarray}\label{z_n^}
z_{\hat n}=\frac{1}{\sqrt{2\hat H+4\alpha-N+1+2\hat F}}\,
\left((2\hat H+2\alpha+\hat F)\,a_{\hat n}-
a^*_{\hat n}\,{\hat a}^2\,\right)~,
\end{eqnarray}
\begin{eqnarray}\label{z_n^*}
z_{\hat n}^* = \left(a_{\hat n}^*(2\hat H+2\alpha+\hat F)-
{{\hat a}^*}\,^2\,a_{\hat n} \right)\,
\frac{1}{\sqrt{2\hat H+4\alpha-N+1+2\hat F}}~,
\end{eqnarray}
where  $\hat F$ is the following real function of scalar operators 
\begin{equation}\label{F}
\hat F =\sqrt{{\hat a}^*\,^2{\hat a}^2+2(2\alpha-N+1)
(\hat H+\alpha)}~.
\end{equation}
By (\ref{M^2^}) $\hat F$ can be written as a function of 
$\hat H$ and $\hat J^2$ and it corresponds to $F_{N+1}$ of \cite{DJ2}.
The operator expressions with square roots in (\ref{z_n^})-(\ref{F})
depend on commuting operators  and there is no ordering problem inside such
expressions. They are naturally  defined on a basis of mutual
eigenstates of $\hat H$, $\hat J^2$ and Cartan generators of the $SO(N+1)$
rotations as multiplication operators.  
The form of the boost operators (\ref{z_n^})  guarantees the correct
commutation relations between the compact and noncompact generators.
The commutators of the operators (\ref{z_n^}) can be calculated by 
the exchange relations between the creation-annihilation and scalar
operators  and these calculations complete the commutation relations of the
$so(2,N+1)$ algebra.

The representations $D_{N+1}(\alpha)$ 
are unitary and irreducible if $\alpha$ is above the unitarity bound
$\alpha>\frac{N-1}{2}$.
The analysis of irreducibility of $D_{N+1}(\alpha)$
uses the relation between the scalar operator $\hat z^2$ and 
${\hat a}^2$ \cite{DJ2}
\begin{equation}\label{z=Fa}
\hat z^2=\hat F\,{\hat a}^2~.
\end{equation}
On the basis of this relation we introduce the quantum analog of 
(\ref{tildez^2}) by
\begin{equation}\label{tildea^2^}
\hat a^2\,|\psi\rangle_{ph}=0~.
\end{equation}
The Hilbert subspace ${\cal H}_{ph}$ defined by this condition 
can be $SO(2,N+1)$ invariant
only if  $\alpha$ is just at this bound 
$\alpha=\frac{N-1}{2}$,
where the representation is still unitary, but becomes reducible. 
At the unitarity bound the operator (\ref{F}) reduces to
$\sqrt{{\hat a}^*\,^2{\hat a}^2}$ and it
vanishes on the solutions of (\ref{tildea^2^}).
Then, the operators (\ref{z_n^}) and (\ref{z_n^*}) on ${\cal H}_{ph}$ become
\begin{eqnarray}\label{z_n^0}
&&z_{\hat n}\,|\psi\rangle_{ph} = \sqrt{2\hat H+N-1}\,\,a_{\hat n}
\,|\psi\rangle_{ph}~,\\
\label{z_n^0*}
&&z_{\hat n}^*\,\, |\psi\rangle_{ph}= 
\left(a_{\hat n}^*\,\sqrt{2\hat H+N-1}\,-
{{\hat a}^*}\,^2\,a_{\hat n}\,\frac{1}{\sqrt{2\hat H+N-1}}\,\right)\,
|\psi\rangle_{ph}~.
\end{eqnarray}
The invariance of ${\cal H}_{ph}$ with respect to the subset of infinitesimal
transformations generated by $E$ and $J_{{\hat m}{\hat n}}$
is apparent. For the boosts $z_{\hat n},~ z_{\hat n}^*$ one has to use  
the commutation relations $[\hat a^2,\,\hat H]=2\hat a^2$, 
$\,[\hat a^2,\,{\hat a}^*\,^2]=4\hat H+2N+2$. Then, for example, from 
(\ref{z_n^0*})
we obtain
\begin{eqnarray}\label{a^2z_n^0}
\hat a^2\,z_{\hat n}^* \,|\psi\rangle_{ph}= 
a_{\hat n}^*\,\sqrt{2\hat H+N+3}\,
\,\hat a^2\,|\psi\rangle_{ph}=0~,
\end{eqnarray}
and altogether find that ${\cal H}_{ph}$ is $SO(2,N+1)$ invariant indeed.

The adjoint form of (\ref{z=Fa}) is
$\,\hat z^*\,^2={\hat a^*}\,^2\,\hat F$ and, therefore,
$\hat z^*\,^2$ vanishes on ${\cal H}_{ph}$ 
\begin{eqnarray}\label{z_n^2}
z_{\hat n}^*\,z_{\hat n}^*\,\, |\psi\rangle_{ph}=0~.
\end{eqnarray}

The obtained  representations of the conformal symmetry  
we denote by ${\cal C}_N$ and now we investigate its structure in more detail.
First note that according to (\ref{z_n^0*}) 
the vacuum is not invariant 
under the action of the operators  $z_{\hat n}^*$ and 
the physical states are obtained by
multiple actions of $z_{\hat n}^*$'s on the vacuum state
$|\psi\rangle_{ph}=
(z_{1}^*)\,^{n_1}\,(z_{2}^*)\,^{n_2}...(z_{N+1}^*)\,^{n_{N+1}}\,|0\rangle$.
To analyze the embedding of the isometry subgroup $SO(2,N)$
in ${\cal C}_N$ we calculate the operator for the Casimir number
of  $SO(2,N)$. From 
(\ref{M_mn^}), (\ref{E^}), (\ref{z_n^0}), (\ref{z_n^0*}) and
(\ref{z_n^2}) we obtain
\begin{eqnarray}\label{C_N}
\left(E^2+\frac{1}{2} J_{mn}J_{mn}-\frac{1}{2}\left(
z_{n}^*\,z_n+z_n\,z_{n}^*\right)\right)\, |\psi\rangle_{ph}= 
\frac{1-N^2}{4}\,\,|\psi\rangle_{ph}~,
\end{eqnarray}
where the index summation goes from $1$ to $N$.
Eq. (\ref{C_N}) indicates that  ${\cal C}_N$ contains the UIR's
of $SO(2,N)$ corresponding to the Weyl invariant mass value only.
Acting on the vacuum state
by the operators $z_{n}^*$ $(n=1,...,N)$ we create a $SO(2,N)$ invariant
subspace ${\cal H}_{-}\subset {\cal H}_{ph}$.
Since the Casimir number and the lowest value of $E$ of this representation
are fixed by $\frac{1-N^2}{4}$ and $\frac{N-1}{2}$, respectively,
we get the representation which is unitary equivalent to $D_N(\frac{N-1}{2})$. 
Similarly, acting by the
operators $z_{n}^*$ $(n=1,...,N)$ on the state $z^*_{N+1}\,|0\rangle$,
we generate another subspace ${\cal H}_{+}$ for a new UIR of $SO(2,N)$, which is unitary
equivalent to $D_N(\frac{N+1}{2})$, since now the lowest
eigenvalue of $E$ is $\frac{N+1}{2}$.  

The direct sum ${\cal H}_{-}\oplus{\cal H}_{+}$ is $SO(2,N+1)$ invariant and
we even have 
\begin{eqnarray}\label{H=H_+H_+}
{\cal H}_{ph}={\cal H}_{-}\oplus{\cal H}_{+}~. 
\end{eqnarray}
To see first the $SO(2,N+1)$ invariance of 
$\,{\cal H}_{-}\oplus{\cal H}_{+}\,$,
it is enough to consider the action of the operator
$z^*_{N+1}$ only.  This operator naturally maps 
$\,{\cal H}_{-}$ to ${\cal H}_{+}\,$, since it commutes with all 
$z^*_{n}$. 
Due to (\ref{z_n^2})  
$z^*_{N+1}z^*_{N+1} |\psi\rangle_{ph}=-z^*_{n}z^*_{n} |\psi\rangle_{ph}$, 
which implies that $z^*_{N+1}$ also maps ${\cal H}_{+}\,$ to $\,{\cal H}_{-}$.
To prove that $\,{\cal H}_{-}\,\oplus\,{\cal H}_{+}\,$ covers all 
${\cal H}_{ph}$, we introduce nonphysical states obtained by 
multiple actions of the operator ${\hat a}^*\,^2$ on the physical
states from  ${\cal H}_{-}\oplus{\cal H}_{+}$. 
The first two levels of the $(N+1)$-dimensional oscillator Fock space contain only the physical states
$|0\rangle$ and $z_{\hat n}^*\,|0\rangle$.
The total number of states on the level $k$ is given by $A_{N+1}^k=\frac{(N+1)\cdot\cdot\cdot(N+k)}{k!}$.
These numbers obey
\begin{eqnarray}\label{A_N^k}
A_{N+1}^k=\left(A_{N}^k+A_{N}^{k-1}\right)+A_{N+1}^{k-2}~.
\end{eqnarray}
$\left(A_{N}^k+A_{N}^{k-1}\right)$ here corresponds to the number 
of physical states on the level $k$ from ${\cal H}_{-}\oplus{\cal H}_{+}$ and
the rest $A_{N+1}^{k-2}$ states are obtained by the action of
${\hat a}^*\,^2$ on all states of the level $(k-2)$.
These nonphysical states are obviously orthogonal to the physical states. 
Thus, the physical and nonphysical states constructed in these
way cover the full Fock space. 
\section{Conclusions}

We investigated the dynamics of scalar massless particles on $AdS_{N+1}$
using the conformal symmetry generated by the $so(2,N+1)$ algebra.
To have invariance under a globally well-defined  conformal group,
$AdS_{N+1}$ has to be mapped to half of an ESU $\rr\times S^N$ and 
extended to the full ESU. Then, what
concerns conformal invariant dynamics, all trajectories are either
completely within the boundary of $AdS_{N+1}$, an equator of the $S^N$, or
have a smooth continuation into the other half of the ESU. 

Hamiltonian reduction leads to a $2N$ dimensional
physical phase space $\Gamma_{ph}$, which is a $SO(2,N+1)$ orbit in the
space of generators of the conformal symmetry and the set of trajectories
on the ESU is mapped one to one to $\Gamma_{ph}/R_+$.

The conformal group $SO(2,N+1)$ can also be considered as the isometry
group of $AdS_{N+2}$, i.e. one dimension higher. 
Then $\Gamma_{ph}$ given by the constraint (\ref{tildez^2}) describes
the kinematical domain corresponding to particle trajectories on the 
boundary of  $AdS_{N+2}$.
This boundary is conformally equivalent to an ESU $\rr\times S^N$, i.e.
the conformal completion of the original $AdS_{N+1}$. 

Quantizing the $SO(2,N+1)$ symmetry we introduced the quantum analog
of the constraint (\ref{tildez^2}) by (\ref{tildea^2^}). Working now with
UIR's $D_{N+1}(\alpha )$ of $SO(2,N+1)$, to guarantee the invariance of the
constraint, one first has to choose $\alpha $ such that the representation
becomes reducible and second to constrain oneselves to the irreducible 
component defined by the constraint.
This fixes the lowest possible energy of the conformally invariant
particle to the unitarity bound of $SO(2,N+1)$, i.e. $\alpha = \frac{N-1}{2}$.
Analyzing the transformation properties
with respect to the isometry subgroup $SO(2,N)$ of the conformal group
$SO(2,N+1)$, we found (indicating by $\sim $ the restriction to
  $AdS_{N+1}$ isometries)
\begin{equation}\label{singleton}
\left [D_{N+1}\left (\alpha = \frac{N-1}{2}\right )\right ]_{\mbox{\scriptsize
    constrained}}~\sim ~D_N\left (\alpha = \frac{N-1}{2}\right )~\oplus
    D_N\left (\alpha =
    \frac{N+1}{2}\right )~.
\end{equation}
This relation is well-known from field theoretical considerations \cite{F2} or
general representation theory \cite{AL}.
The l.h.s. of (\ref{singleton}) is equal to a singleton (Rac) \cite{F1,F2,F3}
and the localization of singletons on the boundary has appeared in field
theoretical terms at various places, see e.g. \cite{sing-bound}.

What we claim to be a new result of this paper is the derivation
completely in terms of particle dynamics. On a more detailed level we should
add: particle dynamics formulated via Hamiltonian reduction in terms of
$(N+1)$-dimensional oscillator variables.

Concerning the isometry group of our original $AdS_{N+1}$ the representations
on the r.h.s. of (\ref{singleton}) have no special properties in comparison to
their relatives at generic values of $\alpha >\frac{N-2}{2} $. As mentioned in
the introduction, their relevance for the massless quantum particle could
have been borrowed from field theory. But then we still would not understand
their necessary combination in the form of a direct sum nor the particles
transformation properties under the larger conformal group. 

There is a very
interesting physical interpretation of (\ref{singleton}): In the generic case
$D_N (\alpha )$ describes a massive (isometric) particle which
lives inside $AdS_{N+1}$, its trajectories never reach the $AdS$-boundary. 
$D_N(\frac{N\pm 1}{2})$ still share the property that the
particle lives in one half of the related ESU. Its classical trajectories
are now null geodesics reaching the boundary and being continued by
reflection. On the other side, the massless (conformal) particle has to
live in the whole ESU. Therefore, it is just the quantum mechanical
superposition of the isometric particle with $\alpha =\frac{N\pm 1}{2}$
living in one or the other half of the ESU.
     
A last comment concerns related issues in field theory \cite{I}.   
There, to handle the problems posed by the lack of global hyperbolicity,
for the Weyl invariant situation three possible quantization schemes
related to transparent and reflective Neumann or Dirichlet boundary conditions
have been considered. Obviously our conformal particle corresponds to the
transparent case, while the both reflective versions are related to the
use of a single $D_N(\frac{N\pm 1}{2})$. That the reflective versions
violate conformal invariance can also be seen directly from
the propagators. They are singular at light cones centered at $AdS$ antipodal
points \cite{I,DSS}, but being antipodal is not preserved under general    
$SO(2,N+1)$ transformations (\ref{AdSconf}). However, note that this objection
applies only if one insists on connecting the notion of masslessness also in
field theory with invariance under a globally well-defined conformal group.

\vspace{1cm}

\noindent
{\bf {\large Acknowledgments}}
\noindent
G.J. is grateful to Humboldt University and AEI, Potsdam for
hospitality. His research was supported by grants from the DFG (436 GEO
17/5/05), GRDF, and GAS. H.D. was supported in part by DFG with the 
grant DO 447-3/3.


\begin{thebibliography}{xx}
\frenchspacing
\bibitem{ADSCFT}
O.~Aharony, S.~S.~Gubser, J.~M.~Maldacena, H.~Ooguri and Y.~Oz,
Phys.\ Rept.\  {\bf 323} (2000) 183
[arXiv:hep-th/9905111].
\\
E.~D'Hoker and D.~Z.~Freedman,
arXiv:hep-th/0201253.
\bibitem{DJ1}
H.~Dorn and G.~Jorjadze,
Fortsch.\ Phys.\  {\bf 53} (2005) 486
[arXiv:hep-th/0502081].
\bibitem{DJ2}
H.~Dorn and G.~Jorjadze,
``Oscillator quantization of the massive scalar particle dynamics on AdS
spacetime,''
arXiv:hep-th/0507031.
\bibitem{F1}
C.~Fronsdal,
Phys.\ Rev.\ D {\bf 10} (1974) 589.
\bibitem{F2}
C.~Fronsdal,
Phys.\ Rev.\ D {\bf 12} (1975) 3819.
\bibitem{F3}
M.~Flato and C.~Fronsdal,
Lett.\ Math.\ Phys.\  {\bf 2} (1978) 421.
\bibitem{EH}
S.W. Hawking, G.F.R. Ellis, {\it The large scale structure of space-time},
Cambridge University Press, 1973
\bibitem{LM}
M.~L\"uscher and G.~Mack,
Commun.\ Math.\ Phys.\  {\bf 41} (1975) 203.
\bibitem{I}
S.~J.~Avis, C.~J.~Isham and D.~Storey,
Phys.\ Rev.\ D {\bf 18} (1978) 3565.
\bibitem{DSS}
H.~Dorn, M.~Salizzoni and C.~Sieg,
JHEP {\bf 0502} (2005) 047
[arXiv:hep-th/0307229].
\bibitem{AL}
E.~Angelopoulos and M.~Laoues,
Rev.\ Math.\ Phys.\  {\bf 10} (1998) 271
[arXiv:hep-th/9806100].
\bibitem{sing-bound}
M.~Flato and C.~Fronsdal,
J.\ Math.\ Phys.\  {\bf 22} (1981) 1100.
\\
A.~Starinets,
Lett.\ Math.\ Phys.\  {\bf 50} (1999) 283
[arXiv:math-ph/9809014].
\\
M.~J.~Duff,
``Lectures on branes, black holes and anti-de Sitter space,''
arXiv:hep-th/9912164.

\end{thebibliography}
\end{document}